\documentclass[aps,prl,twocolumn,groupedaddress,floatfix]{revtex4-1}

\usepackage{graphicx,amsmath,amssymb,amstext,epsfig,color,ulem,bbold,hyperref,soul}
\hypersetup{
    unicode=true,          
    bookmarksnumbered=true,
    colorlinks=true,       
    linkcolor=red,          
    citecolor=[rgb]{0,.7,0}, 
    filecolor=magenta,
    urlcolor=cyan,
    menucolor=blue,
    baseurl=http://www.arxiv.org/abs/,
    linktocpage=true,
    hyperfootnotes=true}
\newcommand{\eqn}[2] {\begin{equation}\label{#1}{#2}\end{equation}}
\newcommand{\eno}[1] {\eqref{#1}}
\newcommand{\gsim}{\lower.7ex\hbox{$\;\stackrel{\textstyle>}{\sim}\;$}}
\newcommand{\lsim}{\lower.7ex\hbox{$\;\stackrel{\textstyle<}{\sim}\;$}}
\def\tauF{{\tau_{\rm form}}}

\begin{document}

\title{Holographic Three-Jet Events in Strongly Coupled $\mathcal{N}=4$ Yang-Mills Plasma}

\author{Jorge Casalderrey-Solana}
\email{jorge.casalderreysolana@physics.ox.ac.uk}
\affiliation{Rudolf Peierls Centre for Theoretical Physics, University of Oxford, 1 Keble Road, Oxford OX1 3NP, United Kingdom}

\author{Andrej Ficnar}
\email{andrej.ficnar@physics.ox.ac.uk}
\affiliation{Rudolf Peierls Centre for Theoretical Physics, University of Oxford, 1 Keble Road, Oxford OX1 3NP, United Kingdom}

\date{December 2015}

\begin{abstract}
We analyse classical string configurations with non-trivial transverse dynamics in $AdS_5$-Schwarzschild.  These strings develop kink-like structures which, via the gauge/gravity duality, can be interpreted as the propagation of hard gluons produced in association with a quark-antiquark pair in a strongly coupled $\mathcal{N}=4$ SYM plasma. We observe the appearance of two physically distinct regimes of the in-plasma dynamics, depending on whether the medium is able to resolve the transverse structure of the string prior to its total quench. From these studies we extract the medium resolution scale of the strongly coupled SYM plasma,  defined as the smallest angular separation between two jets that the medium can resolve, 
$\theta_{\rm res} = \mathcal{C}_{\rm res} ( E /\sqrt{\lambda} T)^{-2/3}$, where $\mathcal{C}_{\rm res}=  \frac{2^{4/3}}{\pi}\frac{\Gamma(3/4)^2}{\Gamma(5/4)^2}$. Our analysis constitutes the first study of proxies for three-jet events in a holographic context.
\end{abstract}


\maketitle


\noindent{\bf 1. Introduction.} 
Three-jet events have a prominent role in our understanding of QCD. Since their first observation in the annihilation of $e^+e^-$ pairs into two quark jets and one gluon jet, this type of events have been crucial to prove the existence of gluons, measure their spin, as well as determine $\alpha_s$ (see \cite{Ali:2010tw} for a recent review). Three- and other multi-jet events are also copiously produced in high energy hadronic collisions, including proton-proton and Pb-Pb collisions at the LHC.

In the latter type of collisions, jets are an important diagnostic tool \cite{Gyulassy:2003mc,Kovner:2003zj,Majumder:2010qh,Mehtar-Tani:2013pia,Qin:2015srf}. In the aftermath of those collisions, both at RHIC and the LHC, small amounts of quark-gluon plasma are formed \cite{Adcox:2004mh,Arsene:2004fa,Back:2004je,Adams:2005dq}. The experimental study of this plasma indicates that it may be best understood as a strongly coupled liquid with no quasiparticles. LHC data has also revealed that this plasma interacts significantly with single and di-jet pairs that propagate through it \cite{Aad:2010bu,Chatrchyan:2011sx,Chatrchyan:2012nia,Chatrchyan:2012gt,Chatrchyan:2012gw,Aad:2012vca,Aad:2013sla,Chatrchyan:2013kwa,Abelev:2013kqa,Chatrchyan:2013exa,Chatrchyan:2014ava,Aad:2014wha,Aad:2014bxa,Adam:2015ewa,Adam:2015doa}. First measurements of three-jet events in the plasma, in the form of neighbouring jets, have 
been reported  in \cite{Aad:2015bsa}.

Given the difficulty of understanding the strong coupling limit of QCD, the gauge/gravity duality \cite{Maldacena:1997re} has emerged as a powerful tool to analyse the dynamics of strongly coupled non-abelian plasmas \cite{CasalderreySolana:2011us}. In this letter, we employ the duality to simulate three-jet events produced in $\mathcal{N}=4$ SYM plasma at strong coupling. We will use this computation to address a simple question: what is the minimum angular separation between two jets that traverse the plasma simultaneously and are resolved by the medium? This is an important phenomenological question to understand Pb-Pb jet data at the LHC \cite{CasalderreySolana:2012ef}. We will confront our results with analogous computations at weak coupling \cite{MehtarTani:2010ma,MehtarTani:2011tz,MehtarTani:2011gf,CasalderreySolana:2011rz,MehtarTani:2012cy}.


\noindent{\bf 2. Three-jet events in holography.}
In the holographic dual of $\mathcal{N}=4$ SYM at infinite 't Hooft coupling  ($\lambda\rightarrow\infty$), energetic massless quarks are described by classical open strings attached to probe D7 branes which fill the  entire $AdS_5$ bulk spacetime \cite{KarchKatz}. The string endpoints are dual to quark-antiquark pairs on the field theory side. This setup has been used in \cite{Chesler-Jets,Chesler-LightQ} to simulate the dynamics of a dressed light quark-antiquark pair that has undergone a hard scattering by considering an initially pointlike open string created close to the boundary with endpoints that are free to fly apart and fall, as a consequence of the bulk gravitational force.
    
The field theory dual of such falling string configuration corresponds to a localised quark-antiquark pair surrounded by soft gluonic fields, which are represented by parts of the string away from the endpoints. In this paper, we will supplement such string configurations  with  an additional, localised  structure on the string worldsheet. This structure evolves into a kink-like formation in which the string doubles back on itself, a natural holographic proxy for a hard gluon in the dual gauge theory \cite{Gubser-Gluon}. Therefore, this string configuration mimics a three-jet event in the dual theory.
 
To generate such strings configurations, we consider pointlike strings with a non-trivial transverse structure.  Parametrizing the string worldsheet by $\sigma^a=(\tau,\sigma)$, the embedding functions $X^\mu(\sigma^a)$ describe the motion of the string in the bulk spacetime with metric $G_{\mu\nu}$. The gravitational dual of thermal $\mathcal{N}=4$ SYM is $AdS_5$-Schwarzschild, whose metric in Poincar\'{e} patch reads
\eqn{AdSMetric}
{ds^2 = \frac{L^2}{z^2}\left(-f(z)dt^2 + d{\bf x}^2 + \frac{dz^2}{f(z)}\right)\,,}
where $L$ is the $AdS$ radius, the blackening function is $f(z)=1-z^4/z_H^4$, with $z_H$ the position of the horizon, and the boundary is located at $z=0$. The temperature of the dual field theory is given by $T=1/(\pi z_H)$.

The dynamics of the classical string are governed by 
\eqn{BulkEOM}
{\partial_a P_\mu^a - \Gamma_{\mu\lambda}^\kappa \partial_a X^\lambda P_\kappa^a = 0\,,}
where $ \Gamma_{\mu\lambda}^\kappa $  are the bulk Christoffel symbols, and  $P_\mu^a \equiv -\tau_f \sqrt{-h}\, h^{ab}\partial_b X^\nu G_{\mu\nu}$ are the worldsheet currents with $\tau_f\equiv 1/(2\pi \alpha')$. The worldsheet metric $h^{ab}$ satisfies $\sqrt{-\gamma}  \, h_{ab}  = \gamma_{ab}  \, \sqrt{-h}$, where $\gamma_{ab} \equiv \partial_a X^\mu \partial_b X^\nu G_{\mu\nu}$ is the induced metric. 

\begin{figure}
\includegraphics*[width=0.49\textwidth]{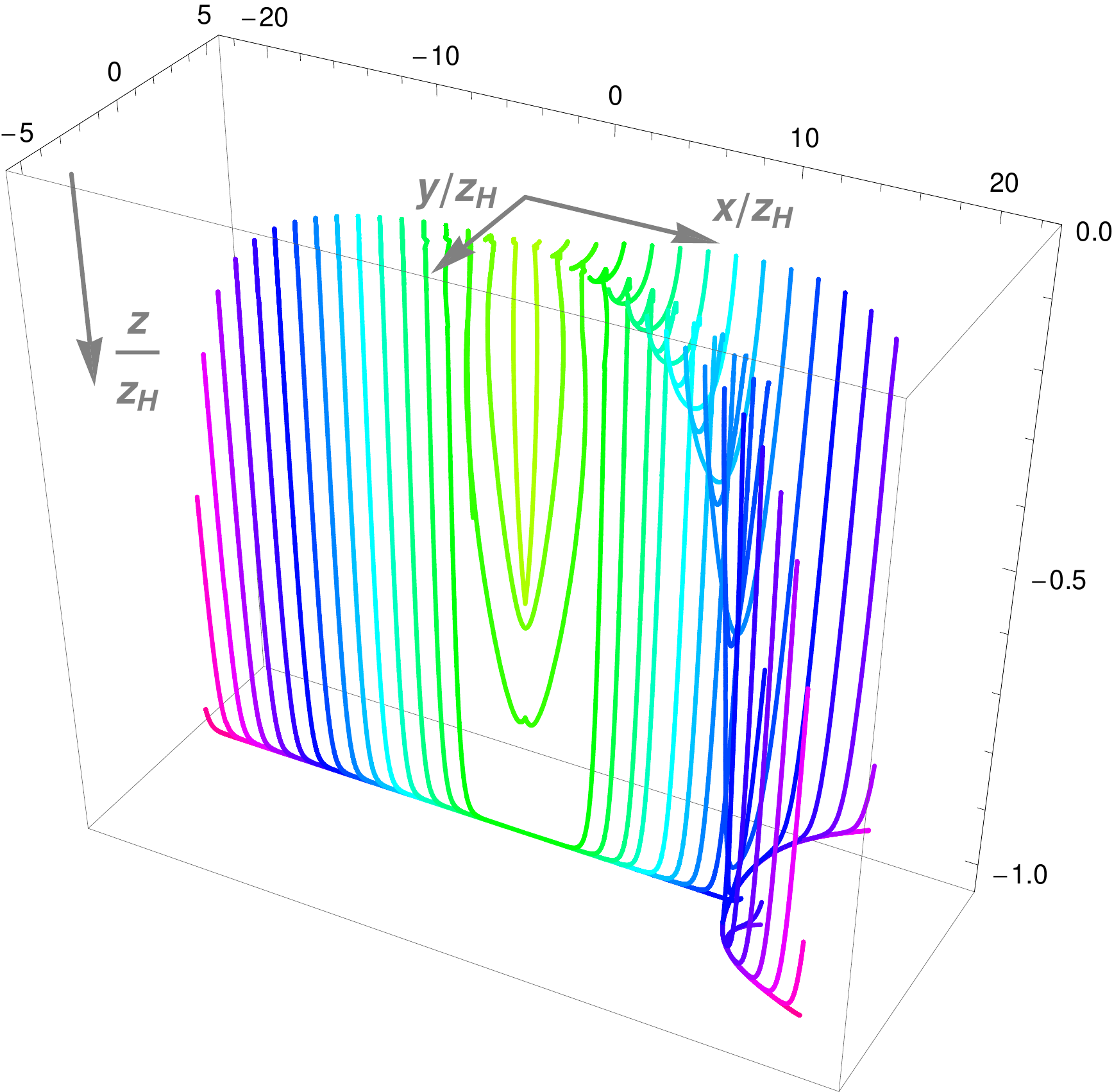}
\caption{Time evolution of a three-jet string with the initial conditions \eno{VelProf}, shown at several fixed times (corresponding to different colors). Parameters used are $\mu_1=0$, $\mu_2=0.6$, $\sigma_1=\sigma_2 =0.2$, $z_0=0.06\,z_H$, $A=4000$ and $B=1432$.}
\label{3DPlot}
\end{figure} 

Pointlike strings are obtained by demanding 
\eqn{PointIC}
{t(0,\sigma)=0\,, \qquad {\bf x}(0,\sigma)=0\,,\qquad z(0,\sigma)=z_0\,,}
where $z_0$ is the radial distance from the boundary at which the string is created, which is dual to the typical size of the excitation in the gauge theory. In the stretching gauge of \cite{Chesler-Jets,Chesler-LightQ}  we choose the initial conditions:
\begin{eqnarray}\nonumber
\dot{x}(0,\sigma)&=&Az_0\cos\sigma, \,\,\,
\dot{z}(0,\sigma)=z_0\sqrt{f(z_0)}(1-\cos(2\sigma)), \\
\label{VelProf}
\dot y(0,\sigma)&=& \frac{Bz_0}{N}\left[Q\,G_{\mu_1,\, \sigma_1}(\sigma-w) -G_{\mu_2,\, \sigma_2}(\sigma-w)\right],
\end{eqnarray}
where $G_{\mu,\, \sigma}(x)=1/\left (\sigma\sqrt{2\pi} \right)\exp\left[-\frac{1}{2}\left(\frac{x-\mu}{\sigma}\right)^2\right] $ is a normalized Gaussian. The forms of  $\dot{x}(0,\sigma)$ and $\dot{z}(0,\sigma)$ are identical to those of  \cite{Chesler-LightQ}, with parameter $A$ controlling the amount of energy in the string. The transverse profile $\dot{y}(0,\sigma)$ is characterised by  parameters $\mu_i$ and $\sigma_i$. The constants  $Q$ and $w$ are chosen such that the total momentum in the $y$ direction vanishes and the boundary conditions are satisfied; $N$ is chosen such that the ratio of the momenta in the $y$ and $x$ directions of the $\sigma =0$ endpoint is $B/A$. Following \cite{Chesler-Jets,Chesler-LightQ}, we numerically solve Eqns. \eno{BulkEOM}, with initial conditions \eno{PointIC} and \eno{VelProf}. In the high energy limit, we observe two distinct behaviours of the string dynamics depending on the values of  $A$, $B$ and $z_0$.

\begin{figure}
\centering
\includegraphics*[width=0.48\textwidth]{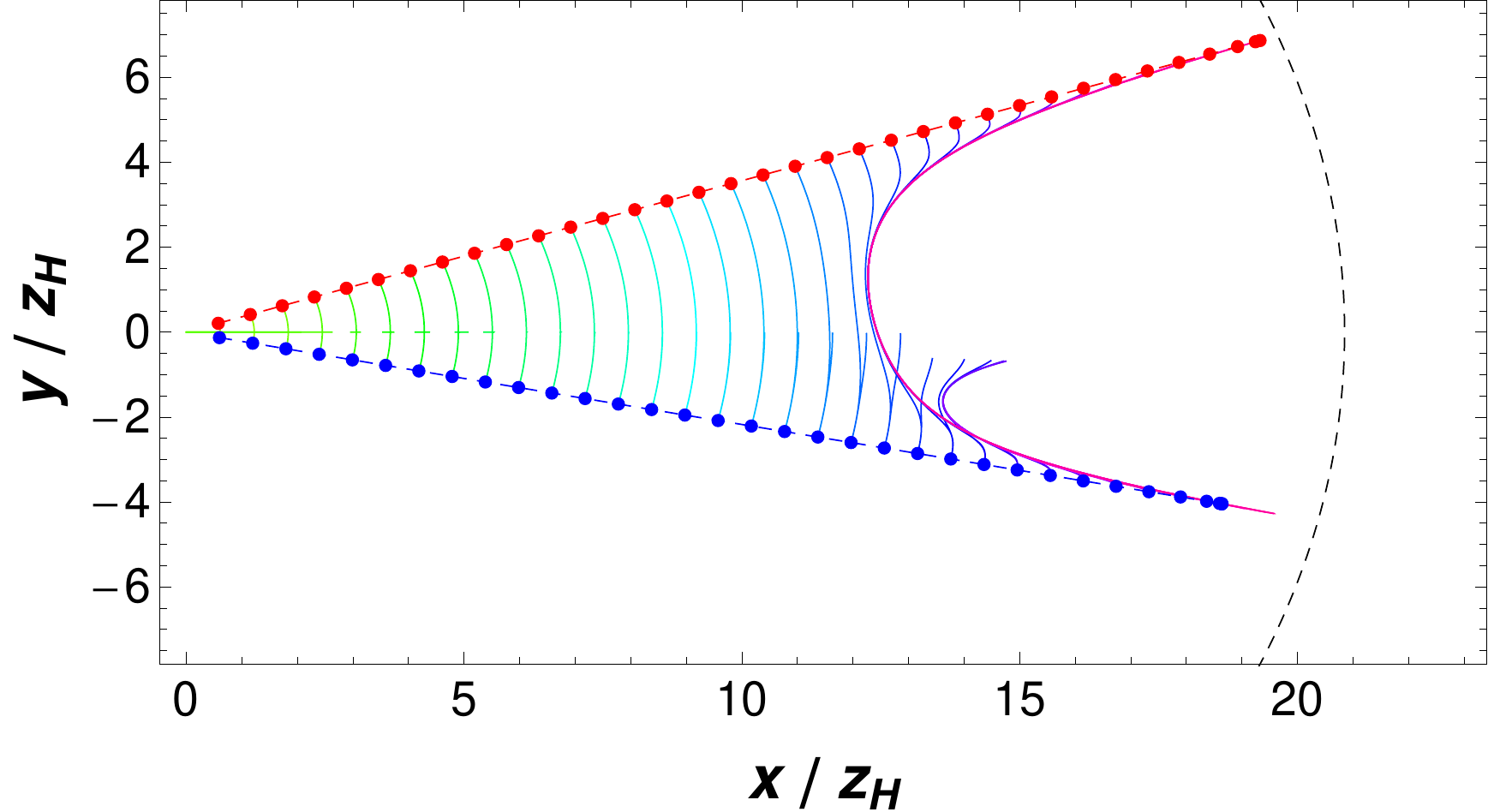}
\includegraphics*[width=0.48\textwidth]{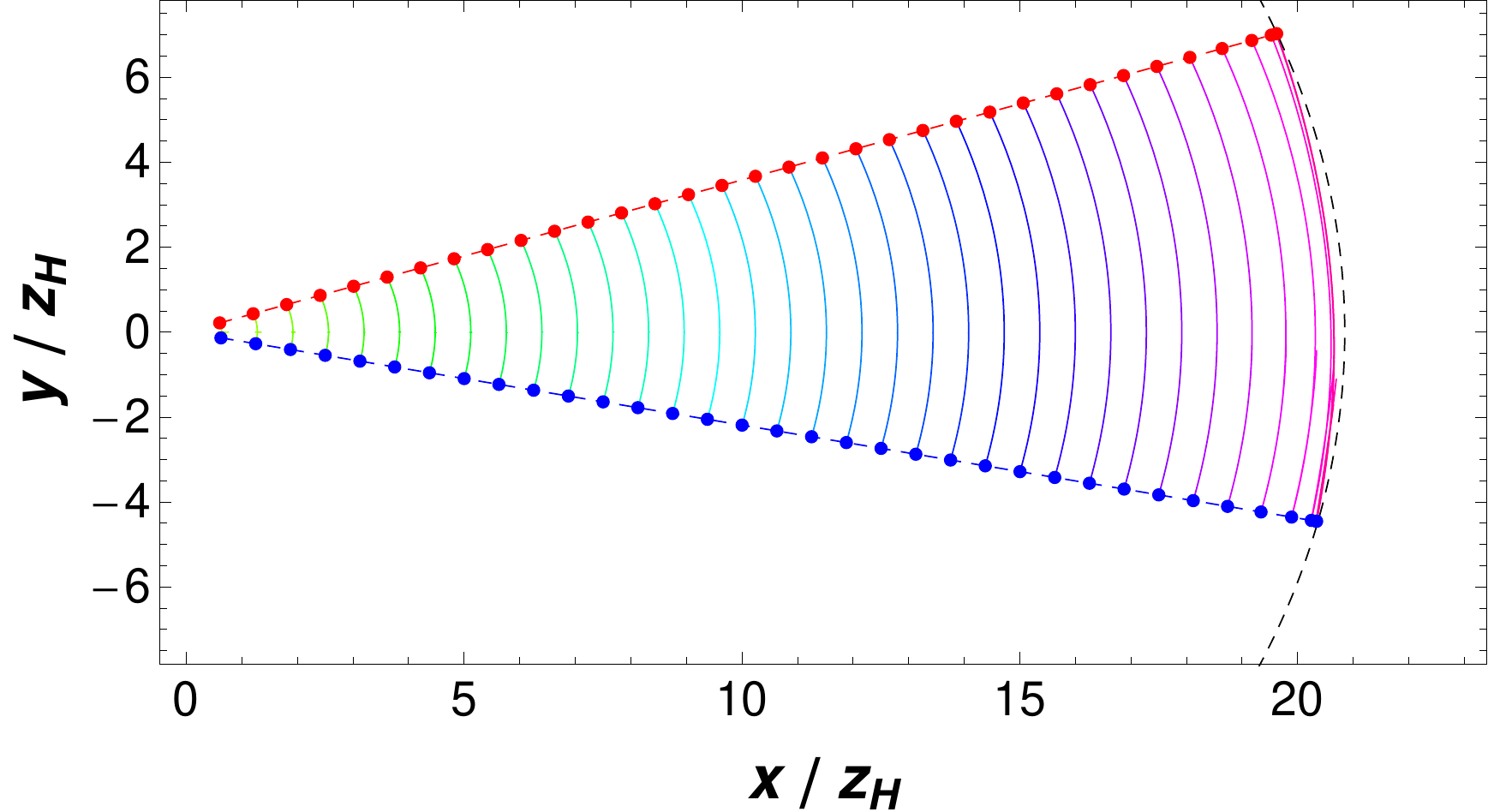}
\caption{
{\it Upper panel:} Time evolution of a three-jet string projected onto the $x-y$ 
plane for the same set of the initial conditions as in Fig. \ref{3DPlot}.
{\it Lower panel:} identical projection, but  with $A=10^5$ and 
$B=35809$. The ratio $B/A$, and hence the angle between the endpoint and the kink $\theta_{qg}$, is the same in both panels. The dashed line indicates the maximum stopping distance, as given by the corresponding null geodesic.}
\label{XYPlot}
\end{figure}

In Fig.~\ref{3DPlot} we show the time evolution of one such  three-jet string. Since our choice of initial conditions \eno{VelProf} constrains the transverse dynamics to the right ($x>0$) direction, the string propagating to the left ($x<0$) is identical to the strings studied in \cite{Chesler-LightQ}.
In contrast, the right-moving string develops a kink-like structure, signaling the formation of a gluon. 
After some time, a part of the string between the endpoint and the kink reaches the horizon while these two points are still at a finite radial distance from the horizon, which allows them to propagate further. In the dual theory, this configuration leads to two well localised excitations disconnected from one another, as clearly seen in the upper panel of Fig. \ref{XYPlot}. We will call this configuration the {\it resolved} case.

The second distinct behaviour is represented in the lower panel of  Fig. \ref{XYPlot}. Here, the ratio $B/A$ and the initial radial depth $z_0$ are kept the same as in Fig. \ref{3DPlot}, corresponding to the same angle between the kink and the endpoint, $\theta_{qg}$, while the energy is increased. In this case, a kink-like structure also forms, but the string connecting the endpoint and the kink falls towards the horizon together with these two points. In the dual theory, the excitation induced by the string in the plasma is broad in the $y$ direction and it does not separate into two well localised excitations. We will refer to this configuration as the {\it unresolved} case. 

\begin{figure}
\includegraphics[width=0.48\textwidth]{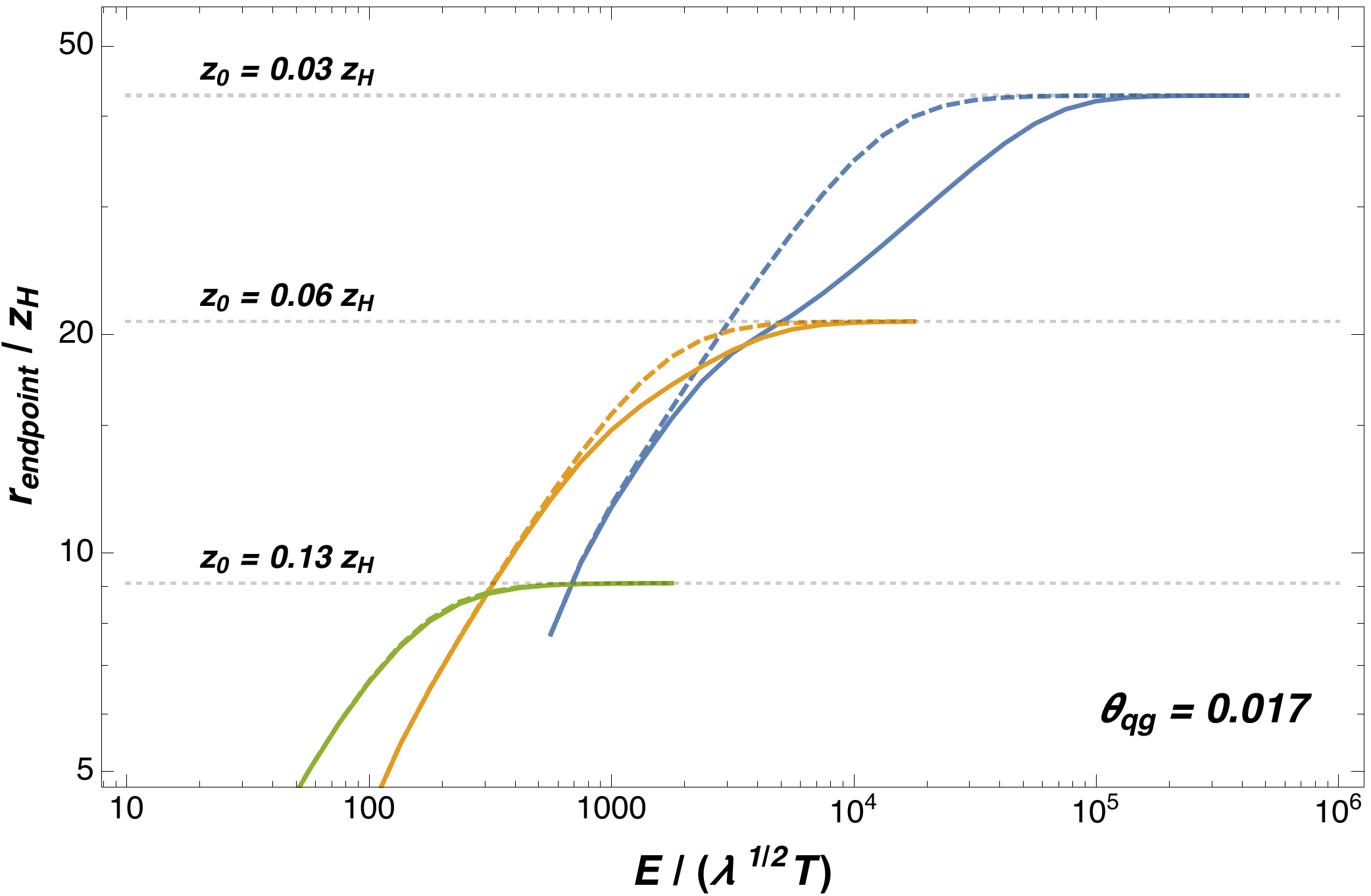}
\caption{Radial stopping distance of the right ($x>0$; solid curves) and the left ($x<0$; dashed curves) endpoint of the three-jet string from Fig. \ref{3DPlot} as a function of energy of the right and left halves of the string, respectively, for a fixed angle $\theta_{qg}$ and several $z_0$. The dashed lines indicate the maximum stopping distances for a particular $z_0$.}
\label{rStopTemp}
\end{figure}

These distinct configurations lead to different energy loss rates in the plasma. For falling strings, such rate is characterised by the maximum stopping distance as a function of the string energy \cite{Chesler:2014jva}. In Fig. \ref{rStopTemp}, we show a comparison of the stopping distances of the  left and right endpoints as functions of the {\it total} energy of the left and right halves of the string, respectively. 
The stopping distance of the left endpoint is identical to that of  the falling strings studied in \cite{Chesler-LightQ}. This is not the case for the right endpoint. At low energies, the stopping distances of both  endpoints exhibit identical energy dependence, indicating that the right part of the string is {\it unresolved} and the dynamics of the right endpoint is independent of the transverse structure. At intermediate energies, and for sufficiently small $z_0$, the relation between the stopping distance of the right endpoint and the string energy depends on the string's transverse profile; these strings correspond to {\it resolved} configurations. At very high energies, the strings again become unresolved as a consequence of the saturation of stopping distances.

\begin{figure}
\centering
\includegraphics*[width=0.48\textwidth]{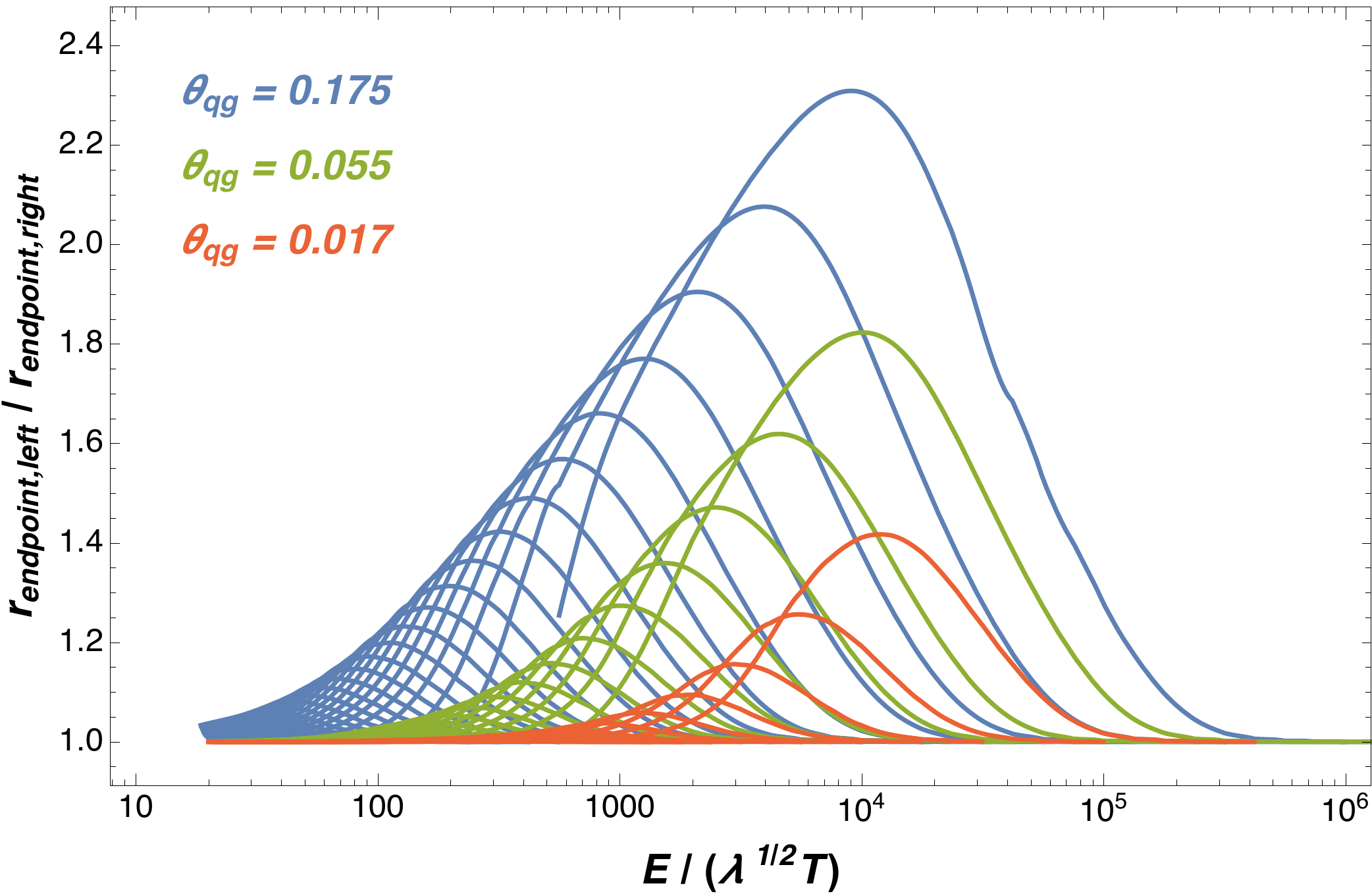}
\caption{Ratio of the radial stopping distances of the left and right endpoints of the three-jet string from Fig. \ref{3DPlot} as a function of energy, for several fixed angles (different colors) and a variety of $z_0$ (different lines).}
\label{FinalRatioPlot}
\end{figure}


\noindent{\bf 3. The medium resolution scale.}  
Based on the previous discussion, we may look for resolved and unresolved cases by computing the ratios of the stopping distances of the two endpoints with respect to the corresponding energy for different $\theta_{qg}$ and different $z_0$. This is shown in Fig. \ref{FinalRatioPlot}, where we see that, for a fixed angle, at low enough energies the system is unresolved. In this way, we may define the medium resolution scale, $\theta_{\rm res}$, as the {\it smallest} angular size of a quark-gluon system with a given energy that a non-abelian plasma of infinite extent can resolve.

For high energy jets $\theta_{\rm res}$  may be determined from  Eqs. \eno{BulkEOM}. We choose  the gauge $\tau=t$ and $\sigma=\theta$, the angle in the $x-y$ plane. We will focus on strings with $z_0\ll z_H$,  which propagate for a long time prior to falling into the horizon, in order to emulate production of three-jet events by processes much harder than $T$. This implies that the string stays at $z\sim z_0$ for a long time, which allows us to approximate the blackening function $f(z)\approx 1$ for most of the propagation of the string. In the high energy limit of large $A$ and $B$ from \eno{VelProf}, most of the energy is assigned to the motion in the $x$ and $y$ directions, resulting in a radial expansion of the string in the $x-y$ plane approximately at the speed of light,
\eqn{xylightrays}
{x(t,\, \theta)\approx t \cos \theta\, , \quad \quad y(t,\, \theta)\approx t \sin \theta\,,}
while, during this time, the string motion in the $z$ direction is subleading, in the sense that $z'\ll x', \, y'$ \footnote{This is valid in regions where the string profile is smooth in $\theta$, as is the case away from the endpoint and the kink.}, where $' \, \equiv{\partial_\theta}$. With these assumptions, the relevant worldsheet currents in the $h_{ab}=\gamma_{ab}$ gauge read
\eqn{NGEnDens}
{\varepsilon \equiv -P_t^t\approx \tau_f \frac{L^4}{z^4}\frac{t^2}{\sqrt{-\gamma}}\,,\qquad 
P_z^t\approx \tau_f\frac{L^4}{z^4}\frac{t^2\dot z}{\sqrt{-\gamma}}\,,}
with $\varepsilon$ approximately constant, as a consequence of Eq. \eno{BulkEOM}. The only non-trivial equation of motion \eno{BulkEOM} is
\eqn{ODE}
{\ddot{z} = 2\left(\frac{\tau_f L^2}{\varepsilon(\theta)}\right)^2\frac{t^2}{z^5} + 2\frac{z^3}{z_H^4}\,,}
where we have again assumed $z\ll z_H$. The first term on the right hand side of \eno{ODE} is the result in $AdS_5$, and the second term is the leading small-$z$ correction arising from the non-trivial blackening function $f(z)$. 

For vacuum jets, $z_H\rightarrow \infty$, and \eno{ODE} can be solved via the scaling $z(t,\, \theta)= z_0 \, \mathcal{F}\left(t/\tauF(\theta)\right)$, with $\mathcal{F}$ a numerically determined function. Since the initial conditions \eno{VelProf} lead to $\dot z (0,\theta)=\mathcal{O} (1/A)$, at high energies we have $\dot{\mathcal{F}}(0)=0$. At early times $t\ll \tauF$, $z$ remains constant, independent of the angle; at late times $t\gg \tauF$, $\dot{\mathcal{F}}$ asymptotes to a constant, leading to a fixed velocity in the $z$ direction, $\dot z \to v^{\infty}_z(\theta)$. Since $z$ is dual to the typical size of the excitation in the gauge theory, in this regime the angular size of the excitations induced by each string bit is  $v^{\infty}_z(\theta)$. Both of these scales are determined by the energy per unit angle in the string, $\varepsilon(\theta)$: 
\eqn{ODE2}
{\tauF=\left( 8 \pi z_0 ^3 \, \frac{\varepsilon(\theta)}{\sqrt{\lambda} }\right)^{1/2}\,,
\quad v^{\infty}_z(\theta)=\mathcal{C} \left( 2 \pi z_0 \, \frac{\varepsilon(\theta)}{\sqrt{\lambda}}\right)^{-1/2}}
where $\mathcal{C}\approx1.29$ and where we used $\sqrt{\lambda}=L^2/\alpha'$.

To have a well defined quark-gluon string as in Fig. \ref{3DPlot}, $\varepsilon(\theta)$ needs to have a $U$-shaped form, with a local minimum at some $\theta_{\rm min}$ between the endpoint and the  kink. This ensures that at $t\gg \tauF$, the velocity $v^{\infty}_z(\theta_{\rm min})$ is maximal. Since $v^{\infty}_z(\theta)$ determines the angular size, the requirement that the quark and the gluon excitations are separated  in the gauge theory imposes that the velocity of both  the endpoint  and the kink must be smaller than $\theta_{qg}$. Similarly, imposing $v^{\infty}_z(\theta_{\rm min})\gg \theta_{qg}$ ensures that the excitation associated with the string away from those points does not overshadow the contribution of the localised quark-gluon excitations. This is similar to the contribution of inter-jet soft radiation in QCD.

\begin{figure}
\includegraphics*[width=0.48\textwidth]{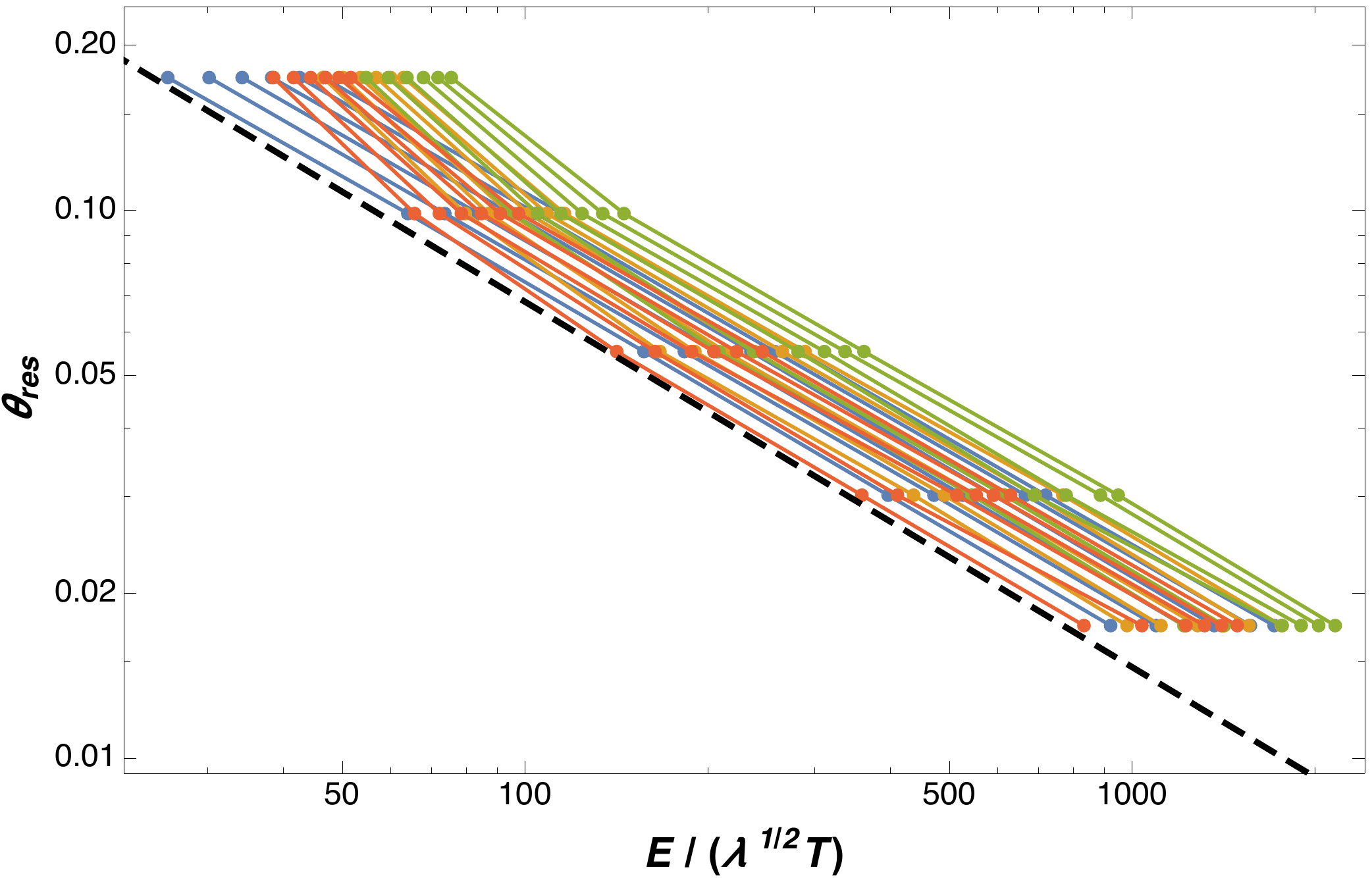}
\caption{The resolution angle $\theta_{\rm res}$ as a function of energy, for several different sets of initial transverse profiles (different colors). Different lines within a given color are obtained by 
extracting the smallest energy at which the ratio from Fig. \ref{FinalRatioPlot} for a fixed angle is equal to 1.05, 1.06, ..., 1.1. Dashed line indicates our analytical result \eno{thetares}.}
\label{VerifyPower}
\end{figure}

In the presence of a black brane, highly energetic strings can fall into the horizon prior to $\tauF$. In those cases, the second term in \eno{ODE}, which is independent of $\varepsilon$, dominates the dynamics. Therefore, the entire string between the endpoint and the kink falls into the horizon at the same time, leading to an unresolved string. Constraining $\tauF > x_{\rm stop}$, with $ x_{\rm stop}=\left(\sqrt{\pi} \Gamma(5/4) / \Gamma(3/4)\right) \, z_H^2/z_0$, the maximal stopping distance of a geodesic initally parallel to the boundary at $z_0\ll z_H$,  the angle of those unresolved strings satisfies
\eqn{thetarunes}
{\theta_{\rm unresolved} <8 \frac{\Gamma(3/4)^2}{\Gamma(5/4)^2} \frac{z_0^5}{z_H^4} \frac{E_{qg}}{\sqrt{\lambda}}
 < 8 \frac{\Gamma(3/4)^2}{\Gamma(5/4)^2} \frac{z_0^5}{z_H^4} \frac{E}{\sqrt{\lambda}} }
where $E_{qg}$ is the energy in the string between the endpoint and the kink, which is always smaller than  $E$, the energy of the entire right half of the string. Figs.~\ref{rStopTemp} and \ref{FinalRatioPlot} show that the resolution angle depends on $z_0$; the smallest possible angle that a {\it resolved} quark-gluon system of energy $E$ can have corresponds to $z_{0*}^3= \sqrt{\lambda}z_H^2/(2\pi E)$,  the value at which the energy dependent stopping distance of a jet saturates its maximal geometric limit \cite{Ficnar:2013wba}. Therefore, the jet resolution angle is
\eqn{thetares}
{\theta_{\rm res} = \frac{2^{4/3}}{\pi}\frac{\Gamma(3/4)^2}{\Gamma(5/4)^2} \left(\frac{E}{\sqrt{\lambda} T}\right)^{-2/3} \, .}
In Fig.~\ref{VerifyPower} we confront this result with the resolution angles extracted from numerical simulations of the three-jet strings with several different sets of initial conditions.

The resolution angle of strings with $\tauF\ll x_{\rm stop}$ is also controlled by $\theta_{\rm res}$.  
In this case, the second term in \eno{ODE} becomes important after the string reaches its asymptotic $v^{\infty}_z(\theta)$ and different string bits possess different stopping distances. These are determined by the maximum distance travelled by a geodesic with initial velocity $v^{\infty}_z(\theta)$. The string will be resolved when the shortest stopping distance of any string bit equals the stopping distance of a jet carrying the entire energy of the quark-gluon system. Since for well defined three-jet strings, $v^{\infty}_z(\theta_{\rm min}) < \theta_{qg}$, these strings will be resolved as long as $\theta_{qg} \gg \,  \theta_{\rm res}$.

\noindent{\bf 4. Discussion.}  
As we have shown, in the infinite coupling limit of $\mathcal{N}=4$ SYM,  $\theta_{\rm res}$ exhibits a characteristic $E^{-2/3}$ scaling with the jet energy. This differs from the scaling expected in perturbative computations. By identifying the coherence length of a dense perturbative  QCD plasma $\tau_{\rm coh}$ \cite{CasalderreySolana:2011rz,MehtarTani:2012cy} with the stopping distance of a jet,  $\Delta x$ \cite{BDMPS}, the perturbative resolution angle scales as $\theta^{\rm pQCD}_{\rm res} \propto E^{-3/4}$ \footnote{Radiative corrections, such as those studied in \cite{Liou:2013qya,Iancu:2014kga,Blaizot:2014bha}, can change this scaling.}. It would be interesting to understand whether the different scaling power is generic of strong coupling by, for example,  exploring different holographic duals. The framework we have developed  can be easily extended to those constructions.

The picture that emerges from this study is much like in pQCD \cite{CasalderreySolana:2012ef}. Even at strong coupling, the interaction of energetic jets in non-abelian plasma may be organised in terms of effective energy loss sources. As we have shown, a coloured excitation of opening angle smaller than $\theta_{\rm res}$ interacts with the strongly coupled medium as a   single coloured object, while well defined three-jet excitations are resolved above this scale. In the latter case, the formation of quark-like and gluon-like strings  observed in Fig.~\ref{3DPlot} suggests that sufficiently resolved multiple jets may lose energy as independent excitations. 

This picture also suggests a possible route to help constrain the dynamics of the QCD plasma from  LHC data. Fluctuations of the energy loss pattern may be used to extract $\theta_{\rm res}$ by correlating these fluctuations with the sub-jet distribution of reconstructed jets or with the number of neighbouring jets as in \cite{Aad:2015bsa}. Further phenomenological studies are needed to gauge the sensitivity of these measurements to this important medium scale.


\noindent{\bf Note added.} While this paper was being finalised, Ref. \cite{Chesler:2015nqz} appeared in which a scaling of the opening angle of jets similar to Eq. \eno{thetares} was found.

\begin{acknowledgments}
We thank P. Chesler, S. Gubser, M. Gyulassy, Y. Mehtar-Tani, J. Noronha, D. Pablos, K. Rajagopal, C. Salgado, A. Starinets, K. Tywoniuk and X.-N. Wang for useful discussions. JCS is a University Research Fellow of the Royal Society and  was supported in part by a Marie Curie Career Integration Grant FP7-PEOPLE-2012-GIG-333786. AF was supported by the European Research Council under the European Union's Seventh Framework Programme (ERC Grant agreement 307955).
\end{acknowledgments}

\bibliography{3Jet}

\end{document}